\documentclass[epj,referee]{svjour}
\usepackage{graphicx}
\begin{document}
\title{EXPERIMENTAL MODELING OF PHYSICAL LAWS}
\author{Igor Grabec}
\institute{Faculty of Mechanical Engineering, University of Ljubljana,\\
A\v{s}ker\v{c}eva 6, PP 394, 1001 Ljubljana, Slovenia,\\ 
Tel: +386 61 1771 605, Fax: +386 61 1253 135, \\
E-mail: igor.grabec@fs.uni-lj.si \\}
\date{Received: date / Revised version: date}

\abstract{A physical law is represented by the probability distribution of a measured variable. The probability density is described by measured data using an estimator whose kernel is  the instrument scattering function. The experimental information and data redundancy are defined in terms of information entropy. The model cost function, comprised of data redundancy and estimation error, is minimized by the creation-annihilation process.
} 
\maketitle
\section{Introduction}
\label{intro}
Quantitative physical explorations of natural phenomena involve three basic tasks: performing experiments, processing data, and modeling physical laws.\cite{fe} The leading trend 
in the development of modern experimental systems is to automatize the first two tasks, while the solution of the critical problem of modeling is still left to intuition. In the recent literature there already appear attempts to program as well the modeling for execution on a computer, especially for data acquisition systems in industrial environments.\cite{gs} Since  measurements are always subject to random  influences,\cite{fe} a statistical approach to modeling is needed. Here we consider the probability distribution as a general basis for modeling of a physical law.  The first step of the modeling is an estimation of probability density function (PDF) from experimental data. The most widely applicable is non-parametric estimation as it requires no a priori assumptions about PDF.\cite{gs,dh} 

From the estimated PDF the experimental physical law can be extracted using the conditional average.\cite{gs} This average represents a non-parametric regression which can be carried out simultaneously with the data acquisition by computer. The structure of the corresponding information processing system resembles a structure of the radial basis function neural network.\cite{gs,cm,gs2} In addition to non-parametric regression, several other paradigms from the fields of artificial neural networks, such as multilayer perceptrons, can be interpreted as automatic modelers of physical laws.\cite{gs,cm,ha} Various algorithms for adapting a selected model to experimental data have already been described,\cite{gs,cm,ha} but the development of fundamental principles for a specification of the model structure is still a subject of current research.\cite{leo} The problems stem from a significant contrast between the complexity of experimental data and the structure of physical laws. The information about the phenomenon explored is generally increased with the number of experimental data; hence instrumental science and technology tend to develop electronic devices with ever greater storage capacity. Contrary to this, the most prominent property of a physical law is its simplicity.\cite{fe} At present it is still not clear how an electronic modeler could automatically and optimally compress the overwhelming experimental data into a simple law, although the theory of algorithmic information has already prepared some fundamentals for the treatment of this problem.\cite{ris,ris2,li} 

A simple model of physical law can be obtained by minimizing a cost function which is composed of model error and complexity.\cite{ris} The theory of statistics offers well elaborated methods for the estimation of the error,\cite{dh,cm,ha} while the description of the model complexity is physically less well established.\cite{gras,ct} For this purpose the measure of algorithmic complexity is applicable,\cite{leo,ris,ris2} but this measure is derived from the program code that determines the average model performance. In the physical literature the complexity is usually considered as an intrinsic property of the phenomenon and should therefore be expressed directly in terms of measured values.\cite{gras} With this aim we define in the next section the experimental information provided by measurements with an instrument of limited accuracy. It turns out that experimental information is useful for the description of the excessive complexity of data which can be utilized for the introduction of the model cost function. 

In order to avoid problems with joining the error and complexity of the model in the cost function, it is convenient to express both terms by a single quantity.\cite{ris} For this purpose we employ the entropy of information,\cite{ct,sha} since it is non-dimensional and provides a common basis for formulation of error and complexity.

\section{Experimental information and redundancy of data}
\label{section1} 
At the definition of the experimental information we consider a scalar-valued variable $X$ since the generalization to a multivariate case is straightforward. For this variable we select a bounded continuous sample space $S_X=(-L,L)$, where $2L$ is the span of the instrument applied. We assume that an arbitrary number, say $N$, of statistically independent measurements has yielded the samples $x_1,\ldots ,x_N$. The non-parametric estimator of PDF is then expressed by the sample average \cite{gs,dh}
\begin{equation}
f(x)\,=\,\frac{1}{N}\,\sum_{n=1}^N\delta(x-x_n).
\end{equation}
This estimator, though unbiased, is not consistent.\cite{gs} As Parzen has shown,\cite{dh,par} it can be made consistent by using as a kernel a smooth approximation of the delta function, such as the Gaussian
\begin{equation}
{\rm g}(x-x_n,\sigma)\,=\frac{1}{\sqrt{2\pi}\,\sigma}\exp \biggl[-%
\frac{(x-x_n)^2}{2\sigma}\biggr].
\end{equation}
with some standard deviation $\sigma$ dependent on $N$. Parzen's estimator 
\begin{equation}\label{pe}
f(x)\,=\,\frac{1}{N}\,\sum_{n=1}^N {\rm g}(x-x_n , \sigma(N)) .
\end{equation}
is therefore biased,\cite{gs} but the bias asymptotically vanishes if $\sigma(N)$ properly
decreases towards $0$ with increasing $N$.\cite{dh,par} The samples $x_1,\ldots ,x_N$ themselves are convenient parameters of the PDF model, but unfortunately their number must increase without limit and the smoothing parameter $\sigma(N)$ is introduced arbitrarily.\cite{cm} Since
measurements are subject to instrumental scattering, the requirement that $\sigma(N)$ vanishes is in conflict with a correct physical presentation of measured quantities.\cite{gs} Consequently, we want to replace Parzen's method by a finite procedure, which would be more in tune with  properties of experiments and would from the very beginning incorporate the measurement inaccuracy in the PDF estimator. 
For this purpose we turn first to the description of the instrumental scattering and interpretation of measured data.

A strict mathematical nalysis of the performance of various PDF estimators has attracted much attention and in more advanced publications on this subject the information entropy is utilized as a common  analytical tool.\cite{cla,haop,hau} However, an exhaustive mathematical analysis of estimator performance apparas too cumbersome for experimentalists which often want to estimate the performance of their estimator already during execution of experiments. Consequently we still utilize the kernel estimator but contrary to Parzen take into account at the description of the kernel the scattering of data caused by measurement procedure and describe the estimator performance by the entrop of information.  

An acqusition of a measured datum can generally be considered as a measurement process in which the measured object generates the instrument output $x$. Common to all meassurements is that there exists an agreement by which the units for the observed variable are selected. Hence we assume that a set of objects which represents the units $\{U_k; k=1,\ldots \}$ is available. Using these objects we can perform a calibration of our instrument. The next common property of measurements is that the outputs of instruments are fluctuating even when callibration is performed. We assume that this property can be characterized by determining the density of the probability distribution of the instrument output at each selected unit. We denote the density of this distribution by $\psi(x|U_k)$. Its mean value $u_k={\rm E}[x|U_k]$ and standard deviation $\sigma$ are usually used to denote the $k$-th element of the scale and scattering of instrument output at the calibration. For the sake of simplicity we further consider the cases where the output scattering does not depend on the position on the scale and can be expressed as a function of $x-u_k$ and $\sigma$ alone: $\psi(x|U_k)=\psi(x-u_k,\sigma)$. Most commonly a Gaussian scattering function $\psi(x-u_k,\sigma)={\rm g}(x-u_k,\sigma)$ is observed. We can generally repeat the calibration procedure with a selected unit $u_k$ finite number of times and obtain a statistical set of calibration samples $\{\psi(x-u_k,\sigma)_n; n=1,\ldots,N_c\}$. If the instrument is well calibated, the scattering functions obtained in repeated callibrations do not differ essentially. The mean of the scattering function over the set of samples is then approximately equal to the result obtained by just one callibration:
\begin{equation}
\psi(x-u_k,\sigma)\,=\,\frac{1}{N_c}\,\sum_{n=1}^{N_c} \psi(x-u_k ,\sigma)_n \,\approx\,\psi(x-u_k ,\sigma)_1,
\end{equation}
In this case we consider the output scattering as a result of inherent fluctuations of measurement procedure and the standard deviation $\sigma$ as the parameter that depends on the quality of the instrument.

Next consider a set of $N$ measurements of variable $X$ with the well calibrated instrument which yield the set of distributions $\{\psi(x-x_i,\sigma);i=1,\ldots,N\}$ with a standard deviation that is practically independent on $i$. In this case we interprete the scattering of mean values $x_1,\ldots ,x_N$ as the consequence of the external variation of the input in repeated mesurements. We therefore consider instrument input $X$ as a random variable and describe its PDF by the mean over the set of experimentally obtained distributions $\{\psi(x-x_i,\sigma); i=1,\ldots,N\}$. The corresponding mixture model \cite{dh}
\begin{equation}\label{mm}
f_N(x)\,=\,\frac{1}{N}\,\sum_{i=1}^N \psi (x-x_i,\sigma) ,
\end{equation}
resembles Parzen's estimator Eq.\,\ref{pe}, but here $\sigma$ is a constant
given by instrument calibration that is independent of $N$. Therefore we also omit in the following text $\sigma$ from $\psi$. If the true probability distribution of variable $X$ is given, then the general properties of this estimator can be analysed following the methods developed by the other authors.\cite{cla,haop,hau}. However, we rather proceed to the definition of experimental information and demonstation of its applicability for the estimation of an optimal number of experiments needed for the specification of the PDF. 
With this aim we first describe the
indeterminacy of variable $X$ in terms of the entropy of information.\cite{ct} For a discrete random variable that assumes $N$ states with probabilities $p_i$
Shannon introduced the entropy of information by \cite{sha} 
\begin{equation}
H=-\sum_{i=1}^N p_i \log p_i .
\end{equation}
It is always between $0$ and log$\,N$ and attains its maximal value when all probabilities are equal: $p_i=1/N$. For a continuous random variable with PDF $f(x)$ the entropy of information must be defined relative to some given reference probability density function $\rho(x)$ as \cite{kol} 
\begin{equation}
H=-\int_{S_X} f(x) \log \frac{f(x)}{\rho(x)} \,dx .
\end{equation}
We will use as the reference the uniform density $\rho(x) = 1/2L$ over the
instrument range for which we get
\begin{equation}
H=-\int_{-L}^Lf(x) \log f(x) \,dx - \log 2L .
\end{equation}
With this formula we first express the uncertainty of the instrument
calibration as 
\begin{equation}
H_u=-\int_{-L}^L\psi(x,u)\log \psi(x,u)\,dx - \log 2L .
\end{equation}
For $\sigma\ll L$ we obtain from the Gaussian scattering function $\psi(x,x_i)={\rm g}(x-x_i,\sigma)$ the approximation
\begin{equation}
H_u\approx \log \frac{\sigma}{L}+\frac{1}{2}\Bigl[1+\log \frac{\pi}{2}\Bigr] ,
\end{equation}
which shows that the uncertainty of calibration depends only upon the ratio
of scattering width $2\sigma$ and the instrument span $2L$. The number
$\log (\sigma / L)$ determines the lowest possible uncertainty of measurement
on the given instrument, as achieved at its calibration.

The indeterminacy of the random variable $X$, which characterizes the scattering of
experimental data, is defined by  
\begin{equation}
H_e=-\int_{-L}^Lf_N(x)\log f_N(x)\,dx - \log 2L
\end{equation}
and is generally greater than the uncertainty of calibration. We define the
experimental information about $X$ by the difference  
\begin{eqnarray}
I=H_e-H_u&=&-\int_{-L}^Lf_N(x) \log f_N(x) \,dx \nonumber \\
&\phantom{=}&+\int_{-L}^L\psi(x,u) \log \psi(x,u) \,dx .
\label{inf}
\end{eqnarray}
For a measurement that yields a single sample $x_1$ the
probability density is given by $f_1(x) = \psi(x,x_1)$, both integrals in
Eq.\,\ref{inf} are equal, and the experimental information $I$ is
zero. For a measurement which yields multiple samples
$x_1, \ldots , x_N$ that are mutually separated by several $\sigma$,
the distributions $\psi(x,x_i)={\rm g}(x-x_i,\sigma)$ are non-overlapping and the
first integral on right of Eq.\,\ref{inf} can be approximated as 
\begin{eqnarray}
-&\frac{1}{N}&\sum_{i=1}^N\int_{-L}^L\psi(x,x_i) \log \Bigl[%
\frac{1}{N}\sum_{i=1}^N\psi(x , x_i)\Bigr] \,dx  \nonumber \\
&\qquad&\approx \log N-\int_{-L}^L\psi(x,x_1) \log \psi(x,x_1) \,dx
\end{eqnarray}
and this yields $I\approx\log N$. When distributions $\psi(x$,$x_i)$ are
overlapping, but not concentrated at a single point, the inequality $0\leq
I\leq \log N$ holds. As the same relation is characteristic of the entropy of
information for a discrete random variable, the experimental information has
a similar meaning to that of the entropy of information for a discrete case. It
describes how much information is provided by a series of $N$ experiments performed by an instrument with the density of scattering distribution $\psi(x,x_i)$. We thus interprete $I$ as a measure of the complexity of experimental data. 

According to the above analysis $N$ repeated experiments can at most provide
$I_{max}= \log N$ of information and this happens when the distributions
$\psi(x , x_i)$ are non-overlapping. Since some overlapping normally takes
place, the actual experimental information $I$ is smaller than $I_{max}$. In
such a case the measurements do not give the maximal possible information, which
means that characterization of the probability distribution by $N$
experiments is to some extent redundant. Accordingly, we define the redundancy
of experimental observation by the difference  
\begin{equation}
R=I_{max}-I .
\end{equation}
This definition is based only on available  experimental data, therefore $R$ can be determined experimentally at each step of data acquisition. It should be pointed out that ou definition differs from the common definition of the redundancy in terms of mutual information which requires specification of joint probability distribution of variables that describe the data samples.\cite{ct,cla,haop,hau} 

If the standard deviation $\sigma$ of scattering is decreased by improving 
the experiment, the redundancy is reduced and tends to $0$ along with
$\sigma$. With an increasing number of samples the overlapping of
distributions $\psi(x , x_i)$ on the average increases and due to this
overlapping $I$ increases more slowly than $I_{max} = \log N$ and tends to a
certain value $I_\infty$ with increasing $N$. Consequently, the redundancy increases 
on the average with the number of samples. Accordingly, the
experimental information $I$ can be interpreted as a characteristic which
determines the number $K$ of non-overlapping distributions that could represent
the experimental observation. This number is defined by 
\begin{equation}
K={\rm e}^I
\end{equation}
and can be determined from experimental data and the scattering function $\psi$.
Asymptotically $K$ tends to a value $K_\infty$, a characteristic, which can
be estimated quite accurately from a finite number of experiments.

\begin{figure}
\centering
\includegraphics[width=3.375in]{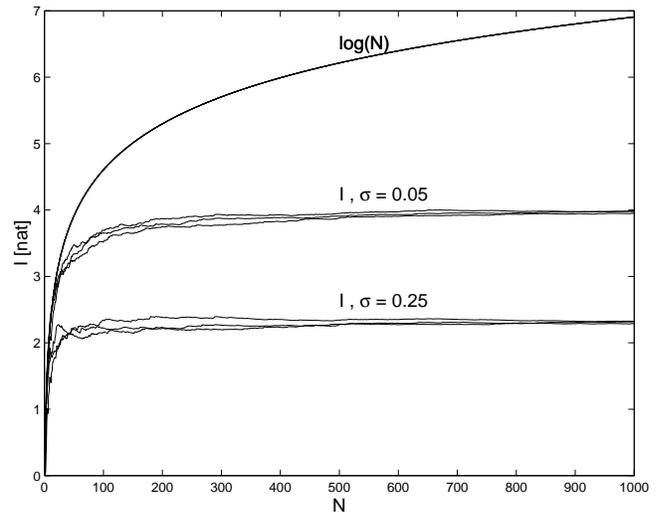}
\caption{Dependence of experimental information $I$ on number of samples
$N$ for a normally distributed variable $X$ with $s=2.5$ and various
instrumental scattering widths $\sigma$.}
\label{figi}
\end{figure}

We illustrate the above--mentioned properties by using a normal random
variable $X$ with standard deviation $s=2.5$. In order to render possible 
a simple setting of its properties in illustrated examples, the samples $x_i$ 
were generated by a computer. Fig.\,\ref{figi} shows the
dependence of the experimental information on the number of samples for two
cases of Gaussian instrument scattering with $\sigma=0.05$ and $0.25$. The
results obtained with three different sample sets demonstrate the
statistical variation of empirical information. In both cases the convergence
of experimental information to a fixed value is observed and the limits
$K_\infty\approx 50$ and $K_\infty\approx 10$ are approximately estimated. As
could be expected, for both cases they are equal to the ratio $s/\sigma$.
Similar results were also observed for the uniform PDF and for mixtures of
normal PDFs. The displacement between the maximal possible experimental
information $I_{max}=\log N$ and other curves in Fig.\,\ref{figi} is the redundancy of
observation. 

\section{Cost function and an optimal number of samples}
\label{section2}
With an increasing number of experimental samples the empirically estimated
PDF converges to a function
\begin{equation}
f_\infty(x) = \lim_{N\rightarrow\infty}
\frac{1}{N}\,\sum_{i=1}^N {\rm g}(x-q_i,\sigma) ,
\end{equation}
which we consider as the hypothetical PDF of variable $X$. Since it can not
be determined by repetition of experiments, we must decide when to stop the
experimentation. From the analysis of the properties of Parzen's estimator
\lbrack 1 - Eq.\,4.19 \rbrack\ we obtain the estimate for the variance
${\rm Var}[f_N(x)]\leq[\sup {\rm g}(x)]^2/N$, which is applicable if the accuracy
of estimation is prescribed. When the accuracy is not prescribed, the
inequality only indicates that $N$ should be increased in order to decrease
the variance; but with increasing $N$ the redundancy increases and we
should consider both properties when deciding about a proper number of
samples $N$. With this aim we utilize two estimators comprising $N$ and $T$
samples. The estimator with $T$ samples is introduced as a reference by which
we estimate the prediction error of the estimator
with $N$ samples. Consequently, $f_T$ should estimate $f_\infty$ with much
greater accuracy than $f_N$ and therefore we take $T\gg N$. We then describe
the estimation error by the Kullback-Leibner information divergence
\cite{gs}
\begin{equation}
D=\int_{-L}^L\Bigl[f_N(x)-f_T(x)\Bigr] \log
\frac{f_N(x)}{f_T(x)} \,dx
\end{equation}
and define the information cost of $f_N$ relative to $f_T$ by 
\begin{equation}
C=D+R_N-R_T .
\end{equation}

\begin{figure}
\centering
\includegraphics[width=3.375in]{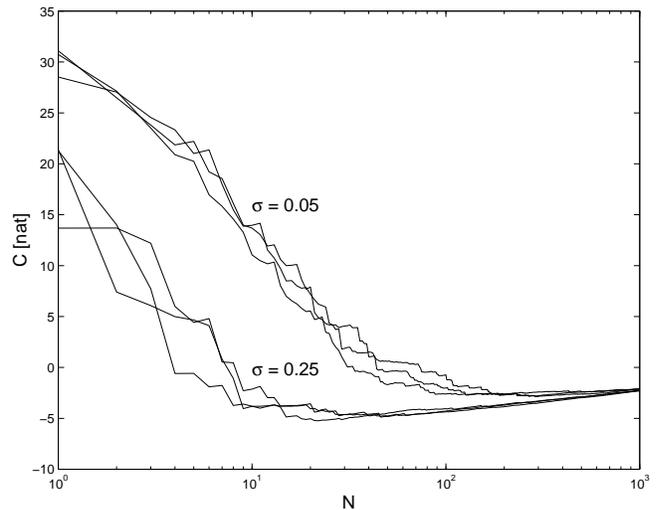}
\caption{Dependence of the cost function $C$ on number of samples $N$ for
a normally distributed variable $X$ with $s=2.5$ and $T=10^4$.}
\label{figii}
\end{figure}

\begin{figure}
\centering
\includegraphics[width=3.375in]{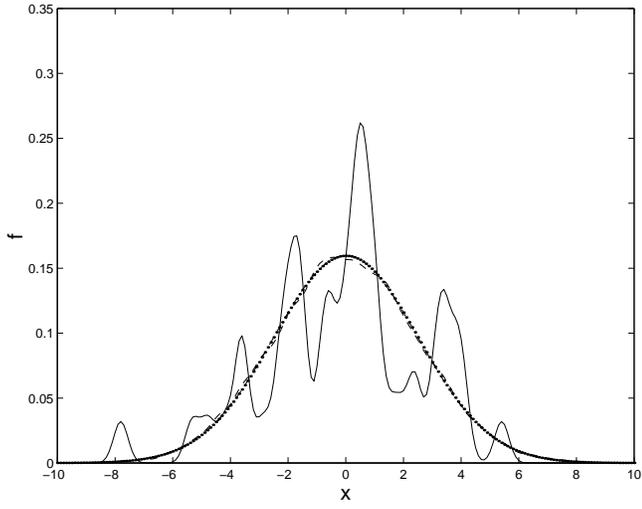}
\caption{Probability density functions $f_{N_o}$ (solid line), $f_T$ 
(dashed line), and $f_\infty$ (dotted line).}
\label{figiii}
\end{figure}

The dependence of $C$ on $N$ with $T=10^4$ is shown in Fig.\,\ref{figii}
for the same data as in the case of Fig.\,\ref{figi}. The number $N_o$ at
which the cost $C$ is minimal is to be considered as the proper number of
samples for modeling of PDF. It depends on the samples used in estimator $f_N$
and we statistically determined $N_o=\,35\,\pm 20$ for $\sigma=0.25$, and
$N_o=\,218\pm 64$ \ for $\sigma=0.05$. The relatively large statistical
scattering of $N_o$ is a consequence of the very slow, approximately
logarithmic divergence of the redundancy. The number $N_o$ also 
depends on the sample set used in estimator $f_T$, but if $T$ is much greater
than $N_o$, its influence is negligible in comparison with statistical
scattering. Fig.\,\ref{figiii} shows an example of the estimated
probability density $f_{N_o}$ for $\sigma=0.25$, $N_o=46$ and
$C(N_o)=-4.8\,$nat. For the purpose of comparison, $f_T$ is 
also shown in Fig.\,\ref{figiii}. Our examples show that the proper
number $N_o$ is several times greater than $K_\infty$. Since $K_\infty$ can
be simply calculated, $N_o$ can be roughly estimated also without calculation
of the cost function.

Figure\,\ref{figiii} shows that $f_{N_o}$ is a rather coarse estimator of probability density. The reason for this property can be explained if the variation of the estimation error and redundancy term in the cost function with increasing $N$ is considered. When $N$ increases the minimum of $C$ is achieved at a low number of samples because of increasing redundancy; hence the estimation error need not be negligible but just properly counter balanced by the redundancy. This further means that a low number of functions ${\rm g}(x-q_i,\sigma)$ with a small $\sigma$ cannot very accurately represent a broad and smooth function $f_T(x)$. 

\section{Generalized PDF model}
\label{section3}
If we want to improve the representation of the PDF by a small number of functions we evidently may not keep $\sigma$ fixed. For this purpose we change the estimator of   
$f_{N_0}$ into a general mixture model 
\begin{equation}\label{gm}
f_M(x)\,=\sum_{i=1}^M \, p_i\,\psi_i(x)
\end{equation}
by using $M$ basis functions $\psi_i(x)=g(x-q_i,\sigma_i)$ and adjustable
parameters $q_i$, $\sigma_i$ and $p_i$. We define here the entropy of basis functions and
the information content of the model as means over probabilities $p_i$  
\begin{eqnarray}
H_s &=& -\int_{-L}^L\sum_{i=1}^M p_i \, \psi_i(x)\, \log 
\psi_i(x) \,\,dx - \log 2L ,\\
I_M&=&H_M-H_s \\
&=& -\int_{-L}^L\sum_{i=1}^M p_i \, \psi_i(x) \, \log
\Bigl[\frac{\sum_{j=1}^Mp_j\,\psi_j(x)}{\psi_i(x)}\Bigr] \,dx .\nonumber
\end{eqnarray}
The model redundancy end estimation error are then 
\begin{eqnarray}
R_M&=&\log M - I_M ,\\ 
D_M&=&\int_{-L}^L\Bigl[f_M(x)-f_T(x)\Bigr] \log
\frac{f_M(x)}{f_T(x)} \,dx .
\end{eqnarray}
With these characteristics we define the information cost of the model relative to experimentally estimated $f_T(x)$ as
\begin{equation}
C_M=D_M+R_M-R_T .
\end{equation}

If we want to adapt the model Eq.\,\ref{gm} to experimental data we must specify the number $M$ and parameters $q_i$, $\sigma_i$, $p_i$ of basis functions.\cite{gs} We cannot achieve this by the variation method since $M$ is an integer number.\cite{dh}. Various methods of growing and pruning have been developed for this purpose in the field of neural networks.\cite{ha,leo,st} The growing methods are mainly utilized when the model is adapted to an increasing number of experimental samples, while pruning is used when a large number of experimental samples is compressed to a smaller number of representative data. In any case a decision about the creation or annihilation of model terms must be reached, based upon some criterion. In the literature various criteria have already been proposed, ranging from purely heueristical to strictly theoretical ones, but at present there is still no generally accepted method.\cite{leo} In our treatment we decide to change the number of basis functions in the model if the cost function $C_M$ is decreased by such action. With this criterion we tested first the annihilation process and then a combined creation-annihilation process, which are described in the following subsections. 

\subsection{Model optimization by annihilation of terms}
\label{section4}
Consider the case when the function $f_T$ is determined by an extensive set of redundant experimental data. We start the adaptation of the model Eq.\,\ref{gm} to these data by  selecting $M=T$ and assigning the values $q_i=x_i$, $\sigma_i=\sigma$, $p_i=1/T$ to parameters of basis functions. After that we consider a model with $M=T-1$ terms. 
If we try to determine the parameters of the compressed model by a strict mathematical procedure based on minimization of the cost function $C_{T-1}$, we obtain a set of non-linear equations that is difficult for further treatment. Less rigorously, but physically more sensibly, we proceed by assuming that an improved model can be obtained by compressing $i$-th and $k$-th term determined by $p_i,q_i,\sigma_i$ and $p_k,q_k,\sigma_k$ into single  $j$-th term with parameters $p_j=p_i+p_k$, $q_j=(p_i q_i +p_k q_k)/p_j$, and $\sigma_j=[\sigma_j^2\,p_i/p_j + \sigma_k^2\,p_k/p_j + (q_i -q_k)^2\, p_i p_k / p_j]^{1/2}$, that represent the common probability, center of gravity and standard deviation, respectively; consequently, the total probability and the first two moments of the probability distribution are preserved. The terms are actually compressed only if the cost function is decreased. In the case of just two terms with equal probabilities and widths it was found numerically that they are compressed only if their centers are separated by less than approximately $3\sigma$. The procedure is then iterated on all terms of the model until all possible compressions are carried out. Fig.\,\ref{figiv} shows a result of this procedure for a bi-modal PDF. From the function $f_T$, determined by $10^4$ experimental data, we obtain, after compression, the model with just two basis functions and significantly reduced redundancy. The agreement between the experimentally estimated $f_T$ and the model function $f_M$ is determined by the prediction error $D_M=0.01\,{\rm nat}$, while  $C_M=0.15\,{\rm nat}$ describes the information cost of such a representation. In this case the cost is mainly determined by the redundancy $R_M=0.14\,{\rm nat}$, which is a consequence of the overlapping of model basis functions.

\subsection{Model adaptation by the creation-annihilation process}
\label{section5}

\begin{figure}
\centering
\includegraphics[width=3.375in]{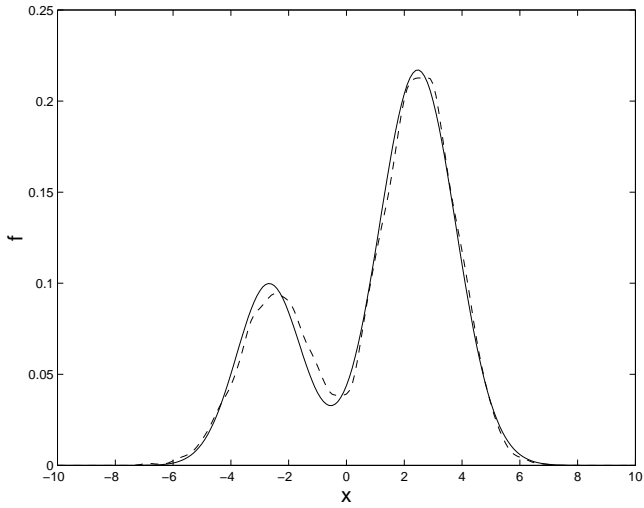}
\caption{Probability density function $f_M$ (solid line) adapted to $f_T$ (dashed line) by the compression of basis functions in the model.}
\label{figiv}
\end{figure}

Although model optimization by annihilation is simple, its weak point is that all the experimental data must be acquired before the start of adaptation. But it is often convenient to form the model simultaneously with acquisition of experimental data. In this case the compression method could still be performed after each acquisition step, but for this purpose all previously acquired data must be stored. We therfore propose a more economical method whereby less numerous model parameters are stored. At $T=1$ we start modeling by setting $f_1 (x)={\rm g}(x-x_1,\sigma)$. After each acquisition step we then create a new term with the parameters $x_T$, $\sigma_T=\sigma$, $p_T=1/T$ and include it in the previous model function by using weighted average $f_T (x)={\rm g}(x-x_T,\sigma)/T + f_{T-1}(x)(T-1)/T$. On this function the compression is then performed. The created term is either annihilated, if the acquired sample $x_T$ falls close to the center of one of the basis functions that comprise the model, or is preserved as an additional term of the model. With increasing $T$ the modification of the PDF by new experimental samples is less and less pronounced. When we perform this procedure with the samples that were used in the preparation of Fig.\,\ref{figiv}, the resulting model PDF agrees with the function, which was obtained by the annihilation process. In the annihilation process the model function is compared with $f_T$ as determined from a large number of samples, while in the creation-annihilation process two successive model functions are compared. Since comparison of model functions can lead to accumulation of errors, one could generally expect smaller modeling error when using the annihilation process. 

\begin{figure}
\centering
\includegraphics[width=3.375in]{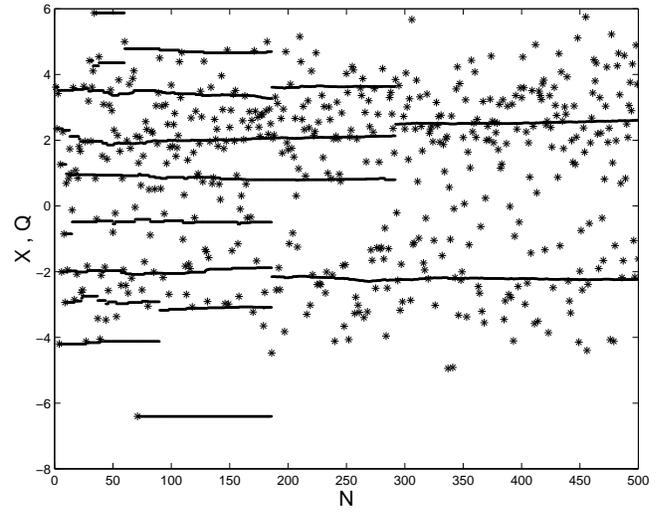}
\caption{Scheme of the creation-annihilation process.
($\ast$ experimental samples, $\bullet$ centers of model basis functions.)}
\label{figv}
\end{figure}

On average the number of model terms in the creation-annihilation process initially increases and subsequently decreases with the number of acquired experimental samples. Therefore, it is instructive to follow the development of the model with an  increasing number of samples. Fig.\,\ref{figv} shows the result obtained during the adaptation of the model to bi-modal PDF of Fig.\,\ref{figiv}. At each acquisition time $T$ the position of the sample $x_T$ is marked by a star, while the centers of basis functions $q_i$ are marked by bullets which may merge into lines. In the initial phase of the model adaptation several basis functions are created and in the later phase some of them are annihilated until ultim•ately an optimal model structure is  established. After that the parameters of the model are less and less influenced by new experimental data. Annihilation of model terms generally keeps the number $M$ of model functions below the number $T$ of samples. Consequently, for large $T$ the storage of model parameters usually requires significantly less memory space than the storage of all the experimental data, and the resulting parameters of the model can often be related to basic processes underlying the investigated phenomenon.

The general mixture model quite often exhibits significantly lower redundancy than the experimental model of Eq.\,\ref{mm}. For example, after compression of the experimental data which determine $f_T$ of Fig.\,\ref{figiii}, we obtained just one term with $q_1$ and $\sigma_1$ determined by the sample mean and standard deviation of variable $X$. These represent a non-redundant optimal model of the hypothetical PDF. A similar conclusion holds for the model of the bi-modal PDF of Fig.\,\ref{figiv}. 

\section{Conclusions}
\label{section6}

We have shown how the PDF of a scalar variable can be estimated non-parametrically by taking into account the inaccuracy of measurements. By the properties of the PDF estimator we have defined the experimental information and redundancy of data. Even though the same definition can be performed with a multivariate variable, the analysis is less comprehensible since the number of parameters in the scattering function increases. We have not specified the form of the scattering function based on fundamental principles, but the central limit theorem of probability indicates that for this purpose a normal distribution could be a proper choice, unless some other is suggested by experiment. The most essential terms of the model cost function are the estimation error and the redundancy. During cost minimization the estimation error provides for a proper adaptation of the model to experimental data, while the redundancy prevents excessive growth of complexity. The search for the cost function minimum yields an estimate of the proper number of the acquisition system data storage cells. The proper number of data cells can be surprisingly low since the redundancy and the divergence are evenhandedly treated in the cost function. If the width of basis functions is determined by experimental scattering only, then the model yields a rather coarse estimate of PDF. The quality of the estimate can often be significantly improved by using the generalized mixture model. The adaptation of the mixture model leads to an effective PDF estimator that is applicable in automatic measurement systems. The creation-annihilation process described also represents a new approach to modeling of artificial neural networks.\cite{gs} In this case the modeler represents a dynamic system with adaptable parameters which are influenced by the experimental data. Evolution of the model terms by creation and annihilation resembles condensation processes in vapors or evolution of grains in alloys and is a typically non-linear, self-organized phenomenon. This analogy indicates the possibility of optimal modeler description by statistical physics and synergetics. 

\section{Acknowledgment}
\label{section7}
The research was supported by the Ministry for Science and Technology of Slovenia and the Volkswagen Foundation in Germany. Prof. T. Klinc from the Faculty of Mechanical Engineering, University of Ljubljana contributed to the preparation of this article by valuable suggestions and critical comments.

\end{document}